\documentclass[a4paper]{article}

\usepackage{icrc2013}

\usepackage[font=footnotesize]{subfig} 

\title{Mirror Development for the Cherenkov Telescope Array}

\shorttitle{Mirror Development for the Cherenkov Telescope Array}

\authors{
A.~F\"orster$^{1}$,
T.~Armstrong$^{2}$,
H.~Baba$^{3}$,
J.~B\"ahr$^{4}$,
A.~Bonardi$^{5}$,
G.~Bonnoli$^{6}$,
P.~Brun$^{7}$,  
R.~Canestrari$^{6}$, 
P.~Chadwick$^{2}$,
M.~Chikawa$^{8}$,
P.-H.~Carton$^{7}$, 
V.~de Souza$^{9}$, 
J.~Dipold$^{9}$,
M.~Doro$^{10}$, 
D.~Durand$^{7}$, 
M.~Dyrda$^{11}$,
E.~Giro$^{12}$,
J.-F.~Glicenstein$^{7}$,  
Y.~Hanabata$^{13}$,
M.~Hayashida$^{8}$,
M.~Hrabovski$^{14}$,
C.~Jeanney$^{7}$, 
M.~Kagaya$^{3}$,
H.~Katagiri$^{3}$,
L.~Lessio$^{12}$,
D.~Mandat$^{14}$, 
M.~Mariotti$^{10}$, 
C.~Medina$^{15}$,  
J.~Micha\l{}owski$^{11}$,
P.~Micolon$^{7}$,
D.~Nakajima$^{16}$, 
J.~Niemiec$^{11}$, 
A.~Nozato$^{13}$,
M.~Palatka$^{14}$,
G.~Pareschi$^{6}$,
M.~Pech$^{14}$,
B.~Peyaud$^{7}$, 
G.~P\"uhlhofer$^{5}$, 
M.~Rataj$^{17}$,
G.~Rodeghiero$^{12}$,
G.~Rojas$^{18}$,
J.~Rousselle$^{19}$,
R.~Sakonaka$^{13}$,
P.~Schovanek$^{14}$,
K.~Seweryn$^{17}$,
C.~Schultz$^{10}$, 
S.~Shu$^{13}$, 
F.~Stinzing$^{20}$, 
M.~Stodulski$^{11}$,
M.~Teshima$^{8,16}$,
P.~Travniczek$^{14}$,
C.~van Eldik$^{20}$,
V.~Vassiliev$^{19}$,
\L.~Wi\'sniewski$^{17}$,
A.~W\"ornlein$^{20}$,
T.~Yoshida$^{3}$,
for the CTA Consortium.
}

\afiliations{
$^{1}$ Max-Planck-Institut f\"ur Kernphysik, Heidelberg, Germany \\
$^{2}$ University of Durham, Durham, Great Britain \\
$^{3}$ Ibaraki University, Japan \\
$^{4}$ Desy Zeuthen, Zeuthen, Germany \\
$^{5}$ IAAT, Universit\"at T\"ubingen, Tuebingen, Germany \\ 
$^{6}$ INAF - Osservatorio Astronomico di Brera, Merate, Italy \\
$^{7}$ CEA, Irfu, Centre de Saclay, Gif-sur-Yvette, France \\
$^{8}$ ICRR University of Tokyo, Japan \\
$^{9}$ Instituto de Física de São Carlos, Universidade de S\~ao Paulo, Brasil \\
$^{10}$ University of Padova and INFN sezione di Padova, Padova, Italy \\
$^{11}$ Institute of Nuclear Physics, Polish Academy of Sciences, Krak\'ow, Poland \\
$^{12}$ INAF - Osservatorio Astronomico di Padova, Padova, Italy \\
$^{13}$ Kinki University, Japan \\
$^{14}$ Institute of Physics of the Academy of Sciences of the Czech Republic, Prague, Czech Republic \\
$^{15}$ Instituto Argentino de Radioastronomia, CCT La Plata-CONICET, Argentina \\
$^{16}$ Max-Planck-Institut f\"ur Physik, M\"unchen, Germany \\
$^{17}$ Space Research Center - Polish Academy of Science, Warsaw, Poland \\
$^{18}$ Universidade Federal de São Carlos \\
$^{19}$ University of California Los Angeles, California, USA \\
$^{20}$ ECAP, Universit\"at Erlangen-N\"urnberg, Erlangen, Germany \\
}

\email{andreas.foerster@mpi-hd.mpg.de}

\abstract{
The Cherenkov Telescope Array (CTA) is a planned observatory
for very-high energy gamma-ray astronomy. It will consist of several 
tens of telescopes of different sizes, with a total mirror area of up 
to 10,000 square meters. Most mirrors of current installations are either
polished glass mirrors or diamond-turned aluminium mirrors, both
labour intensive technologies. For CTA, several new technologies 
for a fast and cost-efficient production of 
light-weight and reliable mirror substrates have been developed
and industrial pre-production has started for most of them.
In addition, new or improved
aluminium-based and dielectric surface coatings have been developed 
to increase the reflectance over the lifetime of the mirrors compared 
to those of current Cherenkov telescope instruments. 
}


\begin{document}
\maketitle


\section{Introduction}
\label{sec:intro}

In recent years, ground-based very-high energy gamma-ray astronomy has
experienced a major breakthrough demonstrated by the impressive
astrophysical results obtained with experiments like H.E.S.S., MAGIC,
and VERITAS~\cite{Aharonian:2008zz}. The Cherenkov Telescope Array (CTA)
project is being designed to provide an increase in sensitivity of at least 
a factor ten compared to current installations, along with a  
significant extension of the observable energy range down to a few 
tens of GeV and up to $>100$~TeV \cite{Hofmann:2010}. 
%
%
To reach the required sensitivity, several tens 
of telescopes will be needed with a combined mirror area of up to 
10,000~m$^2$. 
%
%
Current design
studies investigate three telescope sizes: 
small-sized telescopes with a diameter of approximately 
4~m, medium-sized telescopes (12~m) and large-sized 
telescopes (23~m). In addition, telescopes with dual mirror
optics (Schwarzschild-Couder configuration) are under investigation.

The individual telescopes will have reflectors of up to 400~m$^2$ in area.
The requirements for the point spread function (PSF) are more
relaxed compared to those for optical telescopes. Typically, a PSF below a
few arcmin 
%
%
is acceptable which makes the use
of a segmented reflector consisting of small individual mirror facets 
(called mirrors hereafter) possible. 
Usually, the telescopes are not protected by domes and
the mirrors are permanently exposed to the environment. The design goal
is to develop  low-cost, light-weight, robust and 
reliable mirrors of $1-2$~m$^2$ size with adequate reflectance and focusing
qualities but demanding very little maintenance. 
Current installations mostly use polished glass or diamond-milled aluminium 
mirrors, entailing high cost, considerable time and
labour-intensive machining. 
Most technologies currently under investigation for CTA 
are based on a 
sandwich concept with cold-slumped surfaces made of thin float glass.
In most cases, aluminium honeycomb is used as core material, but
an implementation with v-shaped aluminium spacers exists as well.
In addition, there are sandwich structures made entirely from aluminium
and prototypes made from composite material using a moulding technique
from the car manufacturing industry.


\section{Basic specifications}
\label{sec:specs}

The mirrors for the single-reflector CTA telescopes will be hexagonal in shape,
with sizes of $1-2$~m$^2$, well beyond the
common size of $0.3-1$~m$^2$ of current instruments. 
IACTs are normally placed at altitudes of
$1,000-3,000$~m a.s.l. where significant temperature changes between day 
and night as well as rapid temperature drops are quite
frequent. All optical properties   
should stay within specifications within the range $-15^\circ$~C
to $+25^\circ$~C and the mirrors should resist temperature changes 
from $-25^\circ$~C to
$+40^\circ$~C, with all possible changes of their 
properties being reversible.

Intrinsic aberrations in the Cherenkov
light emitted by atmospheric showers  limit the angular resolution to
around 30 arcsec~\cite{Hofmann:2006wf}. However, the final
requirements for 
the resolution of the reflectors of future CTA telescopes,
 i.e. the spot size of the
reflected light in the focal plane (camera), depend on
the pixel size of the camera and the final design of the telescope
reflector. There is no real need to produce mirrors with a PSF well
below the half of the camera pixel size, which is ordinarily not
smaller than 5 arcmin. A diffuse reflected component is not critical
as long as it is spread out over a large solid angle.
The reflectance into the focal spot 
should exceed 85\% for all
wavelengths in the range from 300 to 550 nm, ideally close to (or even
above) 90\%. The Cherenkov light intensity peaks between 300 and
450~nm, therefore the reflectance of the coating should be optimized
for this range.

\section{Test facilities}
\label{sec:fac}

The standard way to determine the PSF of such mirrors is a so-called 
$2f$-setup: the mirror is placed twice the focal distance $f$ away 
from a pointlike light-source and the return image 
is recorded using CCDs or photodiodes. 
Using waveband filters or narrowband
LEDs, measurements at different wavelengths are possible. Normalizing
for the intensity of the light-source, the total directed reflectance
into the focal spot can be estimated as well. Comparable
setups currently exist in several institutes involved in the development and 
characterization of CTA mirrors. 

While being a reliable method, $2f$-measurements need a lot of space
(several 10s of meters) and are rather time-intensive.
An alternative approach with a compact setup, 
especially for testing large numbers of mirrors,
is being pursued at the University of Erlangen: Phase
Measuring Deflectometry (PMD) \cite{Knauer:2006,Schulz:2011}. 
The basic idea of
PMD is to observe the distortions of a defined pattern after it has been
reflected by the examined surface and from them to calculate the 
exact shape of the surface. For this, sinusoidal patterns
are projected on a screen and cameras take pictures of the
distortions of these patterns.
The primary measurement of PMD is the slope of the mirror 
surface in two perpendicular
directions. A map of the mirror's curvature can be calculated by 
differentiating the slope data. Using a ray-tracing script in
which the normal and slope data from the PMD measurements are the input
parameters, it is possible to calculate the PSF at arbitrary distances from
the mirror.

Cherenkov telescopes usually operate without domes and 
the mirrors are exposed to the
environment for many years. Therefore, an extensive programme of long-term
durability tests is being performed at the University of Durham and the
Max-Planck-Institut f{\"u}r Kernphysik in Heidelberg, trying to use 
ISO standards wherever applicable. Apart from 
classical temperature and humidity cycling for accelerated ageing the 
test series involves 
abrasion tests and sand blasting of mirror surfaces, pull tests
with sticky tape to check the adhesion of the coating, 
tests of the influence of bird faeces on the 
reflective coating, and detailed tests of the water tightness of 
the sandwich structures as well as of their resistance
to mechanical impact from, e.g., hail.

\section{Technologies under investigation for CTA mirrors}
\label{sec:tech}

Several institutes within the CTA consortium have developed
or improved different technologies to build mirrors, 
most of which are in a pre-production phase at the moment.

\subsection{Glass replica mirrors}

\begin{figure}[!t]
  \begin{center}
    \includegraphics[width = 0.8\hsize]{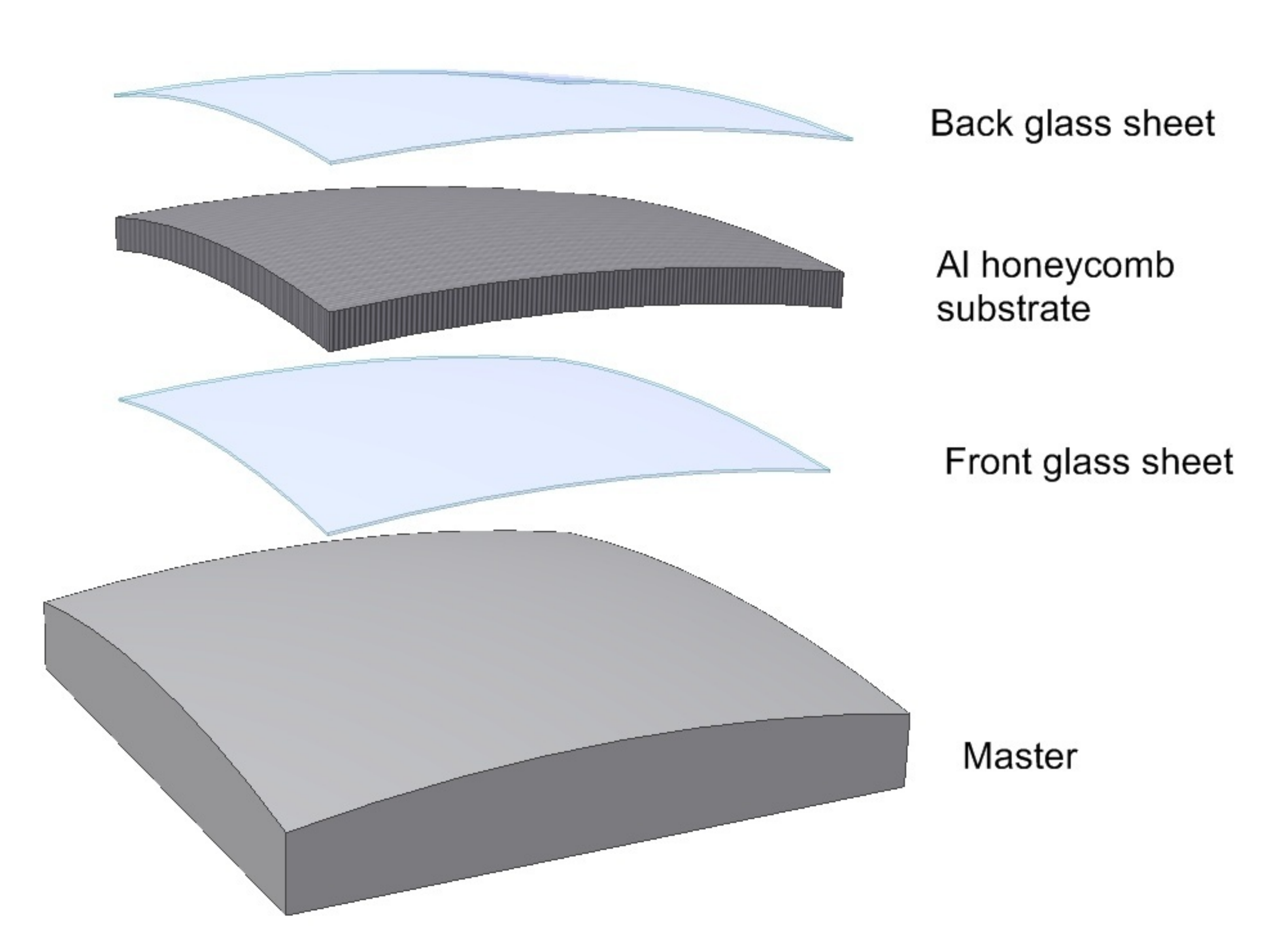} 
    \caption{Cold-slumped glass mirrors (INAF Brera).}
    \label{fig:inaf}
  \end{center}
\end{figure}

The basic concept of this method, originaly developed by INAF Brera,
is to form a thin
sheet of glass on a high precision mould to the required 
shape of the mirror and to glue a structural material and a second glass
sheet to its back to form a rigid sandwich structure.
This concept is being applied by four institutes (INAF Brera, Italy; 
CEA Saclay, France; ICRR, Tokyo, Japan; IFJ-PAN, Krakow, Poland) 
together with industrial partners.
A sketch of the basic layout of this technology is shown 
in Fig.~\ref{fig:inaf}.
\\

\emph{INAF Brera, Italy}

Almost half of the mirror facets of MAGIC II are cold-slumped 
glass-aluminium sandwich 
mirrors~\cite{Pareschi:2008,Vernani:2008,Canestrari:2013}. A thin
sheet of glass is cold-slumped on a high precision spherical
mould. This glass sheet, an aluminium honeycomb and a back sheet are
then glued together with aeronautical glue. The
shaped substrates are coated in the same way as 
traditional glass mirrors. 
A sketch of the design is shown in Fig.~\ref{fig:inaf}.
For CTA the main development goal was to improve the 
process and to reduce the costs.
A first series of 20 mirrors has been produced for the prototype
of the medium size CTA telescope, and these have gone through an extensive
testing programme. 
\\

\emph{CEA Saclay, France}

\begin{figure}[!t]
  \begin{center}
    \includegraphics[width = 0.7\hsize]{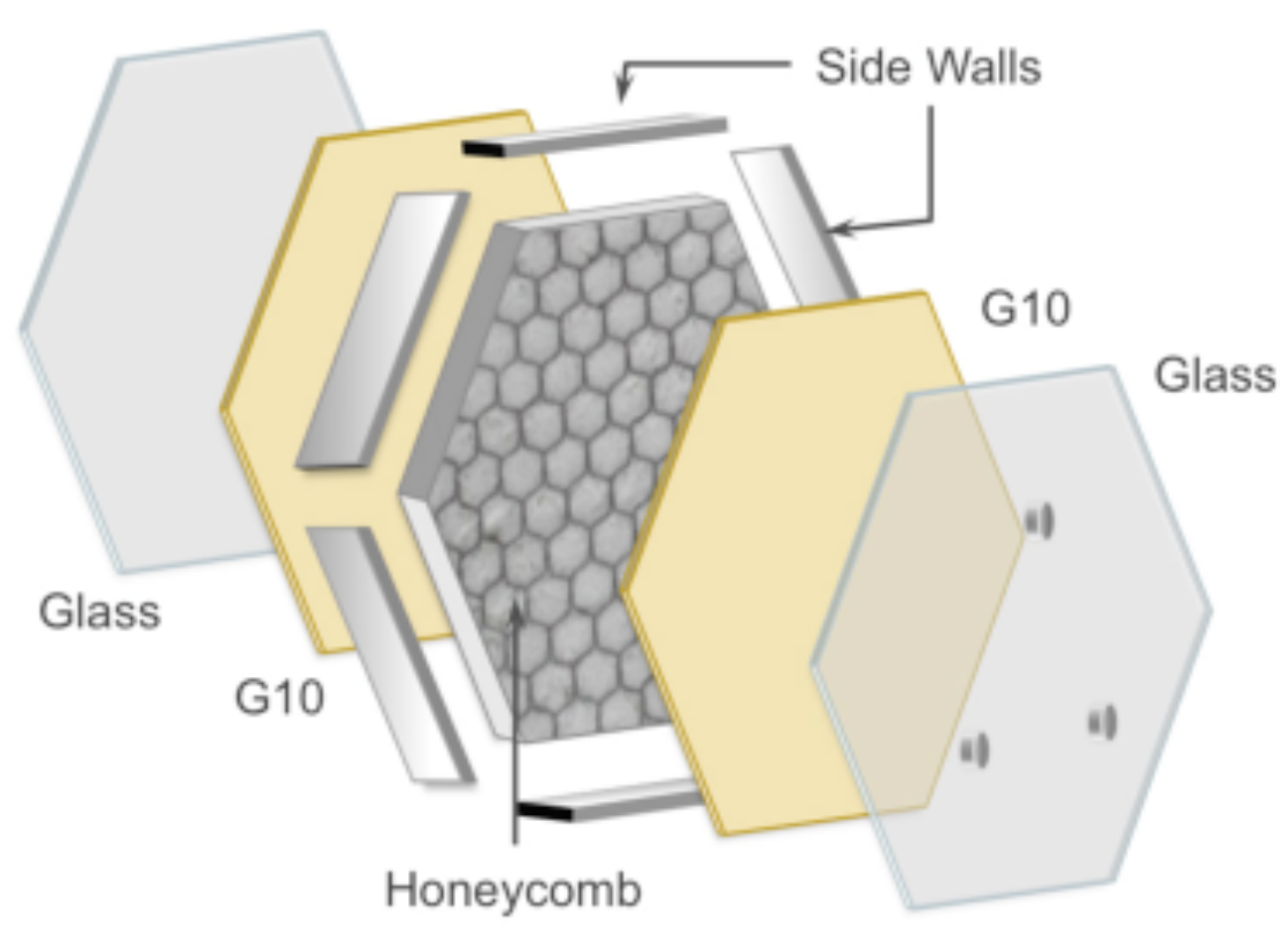} 
    \caption{Cold-slumped glass mirrors with G10 layers (CEA Saclay).}
    \label{fig:saclay}
  \end{center}
\end{figure}

A similar method is being pursued by the Irfu group at CEA (Saclay)
\cite{Brun:2013}, i.e. a sandwich structure is formed by 2 glass sheets
and an aluminium honeycomb core, and the spherical shape of the front surface
is created by cold-slumping the front sheet on a high-precision mould.
Intermediate layers of G10 are inserted between the glass and the honeycomb,
improving the shape and the stiffness of the structure. Rigid side walls are
used to maintain the correct curvature at the periphery of the mirror.
A sketch of the mirror is shown in Fig.~\ref{fig:saclay}.
A series of 30 hexagonal mirrors of 1.2~m flat-to-flat (the planned size for
the medium-size telescopes of CTA) with 16.07~m focal length have been
produced this way. 
\\

\emph{ICRR, Japan}

Slumping technology is also being pursued by the ICRR in 
Tokyo, Japan, concentrating 
on hexagonal mirrors with a size of 1.5~m flat-to-flat as planned for the 
large-sized telescopes of CTA.
The mirrors have a sandwich structure consisting of a glass sheet of 2.7 mm
thickness, an aluminum honeycomb of 60 mm thickness, and another glass
sheet. The reflective layer of the mirror is coated with Cr and
Al on the surface of the glass sheet with a protective multicoat
layer of SiO$_2$, HfO$_2$, and SiO$_2$. A first pre-series mirror production
is ongoing.
\\

\emph{IFJ-PAN, Krakow, Poland}

\begin{figure}[!t]
  \begin{center}
    \includegraphics[width = 1.0\hsize]{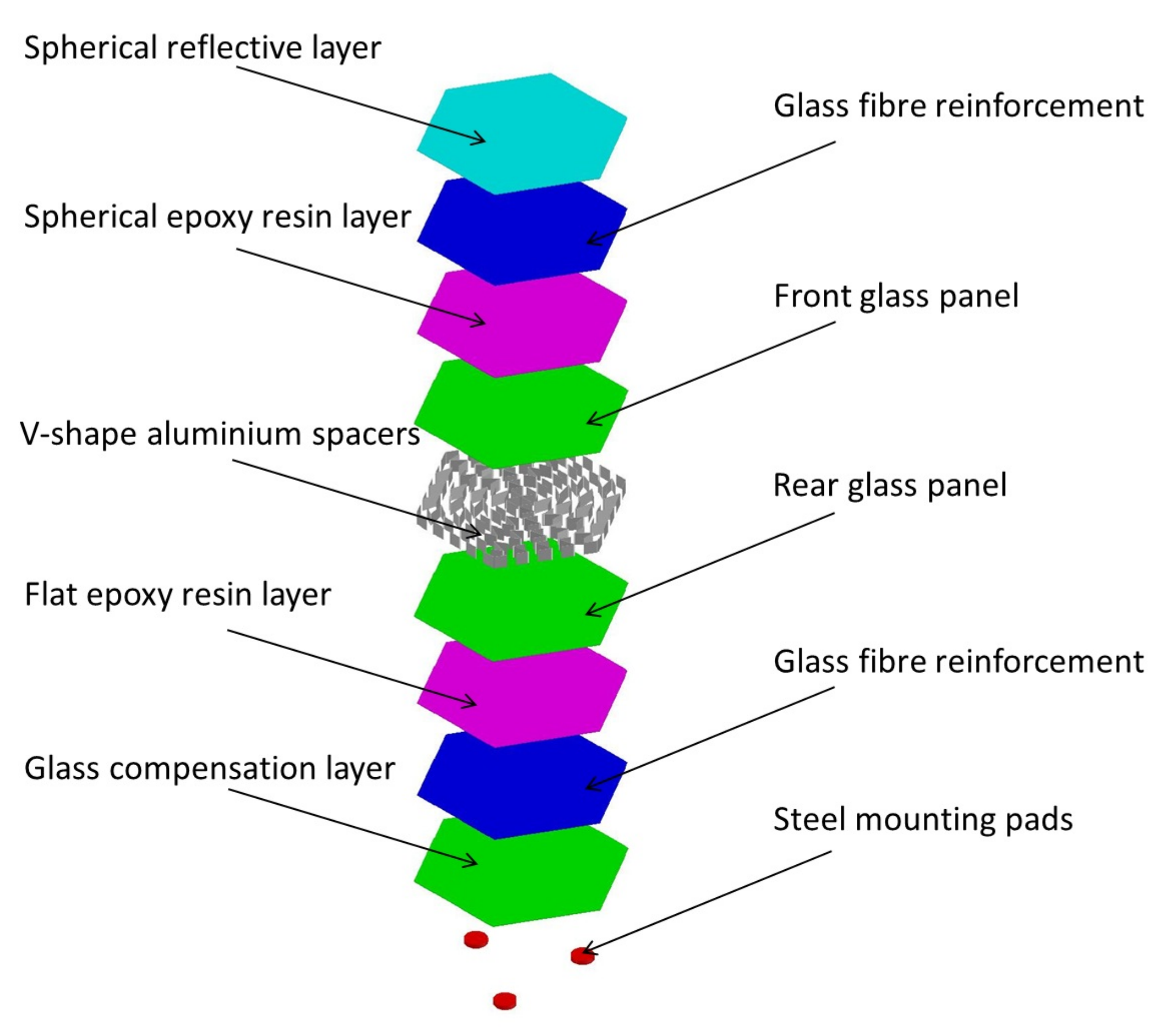} 
    \caption{Cold-slumped open structure glass mirrors (IFJ-PAN Krakow).}
    \label{fig:krakow}
  \end{center}
\end{figure}

The open, cold slumped structure under investigation is a rigid sandwich,
which consists of two flat glass panels separated by v-shaped aluminium
spacers, which are glued using epoxy resin. In a second step an additional,
spherical layer of epoxy resin is formed on the front panel using a master
mould. Then, a reflective layer made of Borofloat 33 glass sheet, coated
with Al and SiO$_2$, is glued to the support structure 
using the same mould. There are also two glass
fibre reinforcements in between the spherical reflective layer/glass
compensation layer and the spherical/flat layer of the epoxy resin to
improve resistance to mechanical impact. 
The open sandwich structure enables good cooling and ventilation of
the mirror panels and avoids trapping water inside the structure. A
stainless steel mesh is attached to the side walls to protect the mirror
structure against contamination by insects or bird waste. A sketch of the
principal design is shown in Fig.~\ref{fig:krakow} and a 
detailed description is given in~\cite{Dyrda:2013}.

\subsection{All-aluminium mirrors} 

\emph{INFN, Padova, Italy}

\begin{figure}[!t]
  \begin{center}
    \includegraphics[width = 0.6\hsize]{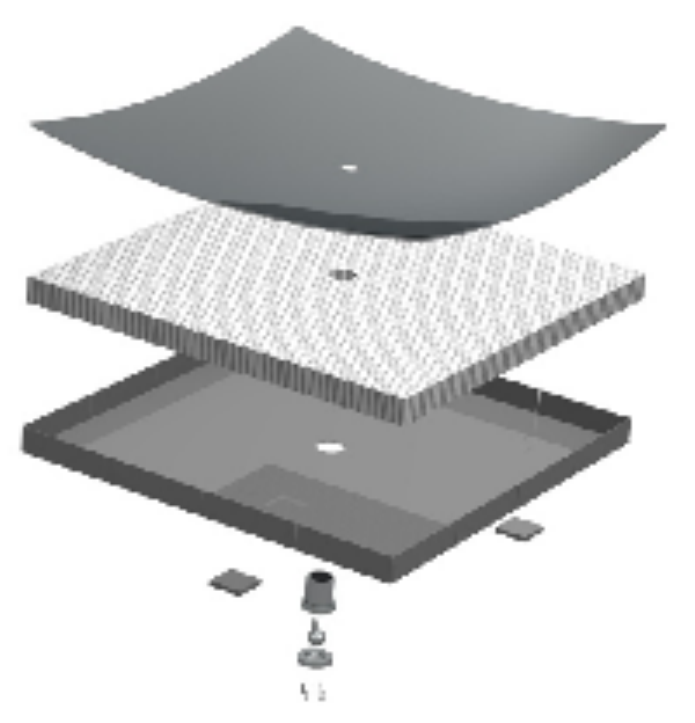} 
    \caption{Aluminium sandwich mirror (INFN Padova).}
    \label{fig:infn}
  \end{center}
\end{figure}

The entire reflector of MAGIC~I and more than half of the MAGIC~II
mirrors are made of a sandwich of two thin aluminium layers interspaced
by an aluminium honeycomb structure that ensures rigidity, high
temperature conductivity and low weight, as shown in 
Fig.~\ref{fig:infn} and described in~\cite{Doro:2008}. 
%
%
The assembly
is then sandwiched between spherical moulds and put in an autoclave,
where a cycle of high temperature and pressure cures the structural
glue. The reflective surface is then generated by precision diamond
milling. 
%
%
The final roughness of the surface is around 4~nm and the
average reflectance is 85\%. The aluminium surface is protected by a thin
layer of quartz (with some admixture of carbon) of around 100~nm
thickness.
%
%
This technology is being further developed for CTA,
particularly by the use of either a thin, coated glass sheet 
or a reflective foil as front layer to reduce the cost imposed by the 
diamond milling of the front surface.
%

\subsection{Composite mirrors}

\emph{SRC-PAS, Warsaw, Poland}

Carbon fibre/epoxy based substrates have good mechanical properties
and show the potential for fast and economical production in large quantities.
The challenge is to produce mirrors with good surface qualities
without labour-intensive polishing. 

\begin{figure}[!t]
  \begin{center}
    \includegraphics[width = 0.5\hsize]{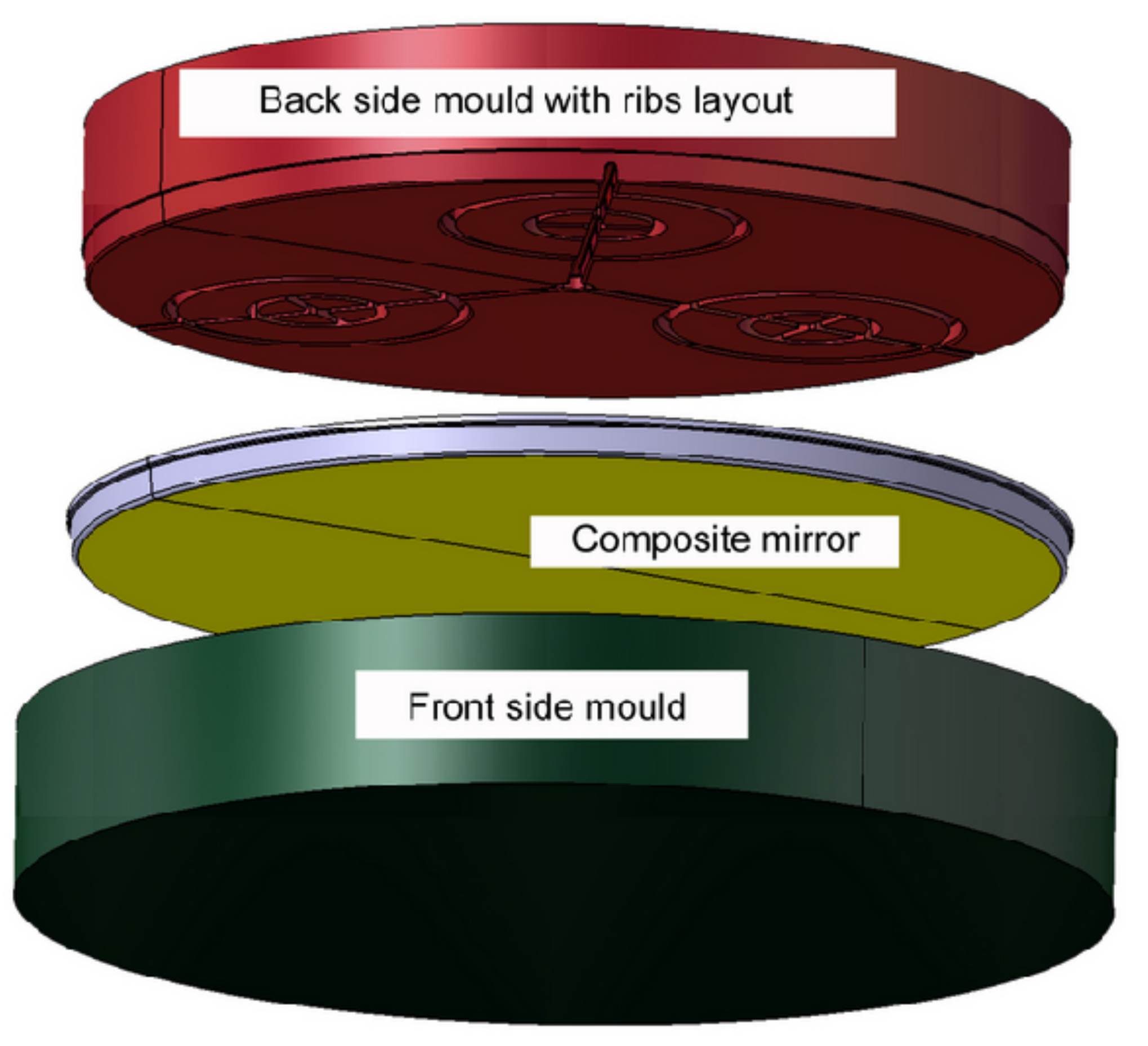} 
    \caption{Composite mirror (SRC-PAS Warsaw).}
    \label{fig:warsaw}
  \end{center}
\end{figure}

Currently, at SRC PAS, the sheet moulding compund (SMC)
technology for composite mirrors is being developed. This
technique has two major features: a) the structure is composed of one
isotropic, thermally conductive material, b) there is no glass in the
structure. The mirror substrate is formed compressing the SMC
material in a mould (see Fig.~\ref{fig:warsaw}). 
The material is a low-cost, widespread and
semi-fabricated product.
SMC is a proven technology in the automotive industry and some of its
advantages include: only a one-step process is needed to produce the substrate,
it is a fast process which takes 3 minutes only and the
material shows no shrinkage.
The top surface of the composite mirror does not require
polishing, as its smooth surface is obtained using a mould 
with a highly polished surface.

\section{Reflective and protective coating}
\label{sec:coat}

Cherenkov telescope mirrors need to have a good reflectance between
300 and 550 nm which makes aluminium the natural choice as
reflective material. 
%
%
%
%
The mirrors are exposed to the environment all year round,
therefore this aluminium coating is usually protected by 
vacuum deposited SiO$_2$ (in the
case of H.E.S.S.), SiO$_2$ with carbon admixtures (for MAGIC) or
Al$_2$O$_3$ obtained by anodizing the reflective Al layer (in the case
of VERITAS). Nevertheless, a slow but 
constant degradation of the reflectance
is observed.

The Max-Planck-Institut f\"ur Kernphysik, Heidelberg,
together with an industrial partner, has performed
studies to enhance both the reflectance and the long-term durability of
mirror surfaces \cite{Foerster:2013}. 
Coatings under investigation include: 
%
%
$a)$ Multilayer dielectric coatings of alternating
layers of materials with low and high refractive index
(e.g. SiO$_2$/HfO$_2$) on top of the aluminization. 
Simple 3-layer designs are already
able to  increase  the reflectance between 300 and 600~nm by 5\%. 
$b)$ Purely dielectric coatings without any metallic layer, which avoids 
the rather low adhesion of aluminium on glass. These show a reflectance 
greater than 95\% in the wavelength region of interest and very low 
reflectance of only a few percent elsewhere. With this the night-sky
background above 550~nm can be significantly suppressed, which is of special
interest if silicon detectors rather than the standard photomultipliers
are used, since they have a higher sensitivity at these wavelengths.
Extensive temperature and 
humidity cycling as well as abrasion tests 
indicate a more stable long-term behaviour 
of these purely dielectric coatings. The disadvantage is that 
mirror with these coatings are more likely to form 
condensation~\cite{Chadwick:2013}. Modifications to improve this
are currently under investigation.

In addition, the University of T{\"u}bingen is working on simulations
to improve the design of the multi-layer coatings and 
operates a coating chamber for the production of 
small mirror samples to study systematically various coating options 
and their production techniques \cite{Bonardi:2011}. 
Furthermore, the University of Sao Paulo is working
to improve the classical Al + SiO$_2$ coating.

%
%

\section{Summary}
\label{sec:sum}

Many of the challenges induced by the demand for a 
few thousand mirrors with a total reflective area
of up to 10,000~m$^2$
for CTA have been technically solved on the prototyping level,
such as the production of large-sized facets of up to 2~m$^2$ in area with low
weight and high optical quality. Currently pre-production series
have been or are being produced for the different technologies
to prove that an easy and rapid series
production at reasonable costs is possible. 
For the mirror coatings,
alternatives to the standard solution have been developed
that show improved durability in laboratory tests. 
For quality control, extensive test facilities have been
set up, for testing both the optical performance of the mirrors and the 
durability of substrates and of coatings.

\section*{Acknowledgements}

We gratefully acknowledge support from the agencies and
organisations listed in this page: 
http://www.cta-observatory.org/?q=node/22.

\end{document}